\documentstyle[11pt,paspconf,epsf]{article}
\begin{document}
\title{On the relative orientation of binary galaxies.}

\author{Iv\^anio Puerari}
\affil{Instituto Nacional de Astrof\'{\i}sica, \'Optica y Electr\'onica, 
Apartado Postal 216, 72000 Puebla, Mexico}

\author{Carlos Garc\'\i a-G\'omez and Antonio Garijo}
\affil{Universitat Rovira i Virgili, Carretera de Salou, s/n,
       43006 Tarragona, Spain}
       
\begin{abstract}
The projected directions of the rotation axes of interacting binary disk 
galaxies tend to align orthogonal to each other. Sofue (1992) has suggested
that this could be due to shorter merger times for galaxies with paralel
spins. We show by means of N-body simulations that this suggestion is correct. 
\end{abstract}

\section{\bf Introduction}

The relative orientation of galaxy spins with the orbital spin in 
interacting binary systems has been examined by Sofue (1992). He found that
the projected directions of the rotation axes tend to aling orthogonal to
each other. He suggested that this could be explained in the case that paired
galaxies with parallel spins have already merged, while those with the
orthogonal spin axes are still in the process. So, systems with a ``Tri-Axial''
angular momentum distribution should have longer lifetimes. This assumes an 
scenario where both galaxies are formed altogether and the plane of the 
orbit lies near the planes of the galaxies. 

It is know that the tidal disturbance on a galactic disc due to a companion 
is strongest in the case of prograde. Thus, binary systems where the 
spins of the galaxies are parallel to the orbital spin would suffer strong 
couplings and will merge very fast (Keel 1991). So, the chance for survival 
of this kind of system would be small. Even when one galaxy has
an anti-parallel spin, the galaxy which have the spin parallel
to the orbital spin will ``see'' the other galaxy
in a direct (prograde) orbit and the chance for an anti-spin
pair to survive a merger is also small. In this scenario,
the ``more stable'' pair, i.e., the system which will take
a longer time to merge is a ``tri-axial'' one, where neither
the galaxies spins nor the orbital spin are aligned.

We will show that the merging time certainly depends on the orientation of 
galaxy spins, but it could also exist a dependence on the energy of the orbit 
and on the time of first pericenter passage. In this work, we perform a 
series of binary disk galaxy interactions to check the validity of the 
``Tri-Axial'' hypothesis.

\section{\bf Simulations}

All the simulations were performed with the TREECODE (Barnes and
Hut 1986). The parameters for the evolution were
a tolerance parameter $\theta$=0.7, a softening equal to 0.06666 and
a time step equal to 0.025. With these parameters the total energy
is conserved better than 0.05~\%. The units of length and time are
3 kpc and 10$^7$ years respectively, and we take $G$=1. Using this 
normalization the unit of mass is $6\times10^{10}~M_{\odot}$.

Each galaxy is formed by an halo and a disk. The spherical system is
a truncated Plummer model while the disk follows a truncated Kuzmin/Toomre law 
in the radial direction and is an isothermal sheet vertically, with 
constant vertical scale thickness. Each model galaxy consists of $40000$ equal 
mass particles and the
halo mass ratio is set to $M_D/M_H = 1/3$. This value and the values of the
Toomre (1964) $Q$ parameter 
are selected in order to prevent the possible formation of bars in the disks
(Athanassoula and Sellwood 1986). Table I shows more details on the parameters 
of each galaxy: $N_H$ and $N_D$ are the number of the particles in
each component, $M_H$ and $M_D$ their masses, $b_H$ and $b_D$ the radial 
scale lengths and $R_{cut_H}$ and $R_{cut_D}$ the truncation radius. For the
case of the disk $z_0$ represents the vertical scale length.

\begin{table}
\caption{Model parameters.}
\begin{tabular}{lllllllllll}\hline
 Halo & $N_H$  &  $M_H$  &  $b_H$  &  $R_{{cut}_{H}}$ &  Disk & $N_D$  &  
 $M_D$  &  $b_D$  &  $R_{{cut}_{D}}$  &  $z_0$ \\
 & 30000  &  1.5  &  5.0  &  10.0 &
       & 10000  &  0.5  &  1.0  &  5.0  &  0.1  \\
\end{tabular}
\end{table}

The binary galaxy systems are set as follows. First, we shifted the isolated 
galaxy model in the $Y$-axis. Galaxy 1 is always placed at $Y=-11$ and Galaxy 2
at $Y=+11$ in simulation units. In this way the halos do not overlap
initially. Their relative speeds are such that the galaxies are initially in
a parabolic orbit with radius of pericenter $R_p=6$ and lie in the $X$-$Y$ plane
(in other words, the orbital spin is always pointing in the $Z$ direction). 
$X_{\pm}Z_+$ means that the spin of the galaxy lies in the 
$X$-$Z$ plane forming an angle of 45$^{\circ}$ ($+$) or 135$^{\circ}$ ($-$) 
with the $X$-axis. In Table II we list the spin orientation of
each galaxy of our simulations.

\begin{table}
\caption{Spin of the galaxies.}
\begin{tabular}{lllllllll}\hline
 & O1 & O2 & O3 & O4 & O5 & O6 & O7 & O8\\\hline
Galaxy 1 & $Z_+$ & $Z_+$ & $Z_+$ & $Z_+$ & $Z_-$ & $X_+Z_+$ & $X_+Z_+$ &$X_-$\\ 
Galaxy 2 & $Z_+$ & $Z_+$ & $Z_-$ & $X_-$ & $X_+$ & $X_-Z_+$ & $X_-Z_+$ &$Y_+$\\
Orbit & $Z_+$ & $Z_-$ & $Z_+$ & $Z_+$ & $Z_+$ & $Z_+$ & $Z_-$ & $Z_+$ \\
\end{tabular}
\end{table}

\section{Preliminar results}

In Figure 1, we plot for all orientations listed in Table 2, the relative 
distance between the mass center of the disks as a function of time, computed 
with the bounded particles. We can see that there is a clear dependence of 
the merging time on the galaxy orientation, in the sense suggested by 
Sofue (1992), the tri-axial systems lasting longer time. For the simulations
O1, O3, and O8, we also performed parabolic 
encounters with pericenter $R_p=12$ and also circular encounters. In the case 
of parabolic orbits with the greatest pericenters the galaxies did not merge 
within 15 Gy. In the case of circular orbits, both galaxies merge as in the
parabolic encounters with small pericenter. There are only small 
differences of the merging times with respect to the case of parabolic 
encounters.

\begin{figure}
\includegraphics{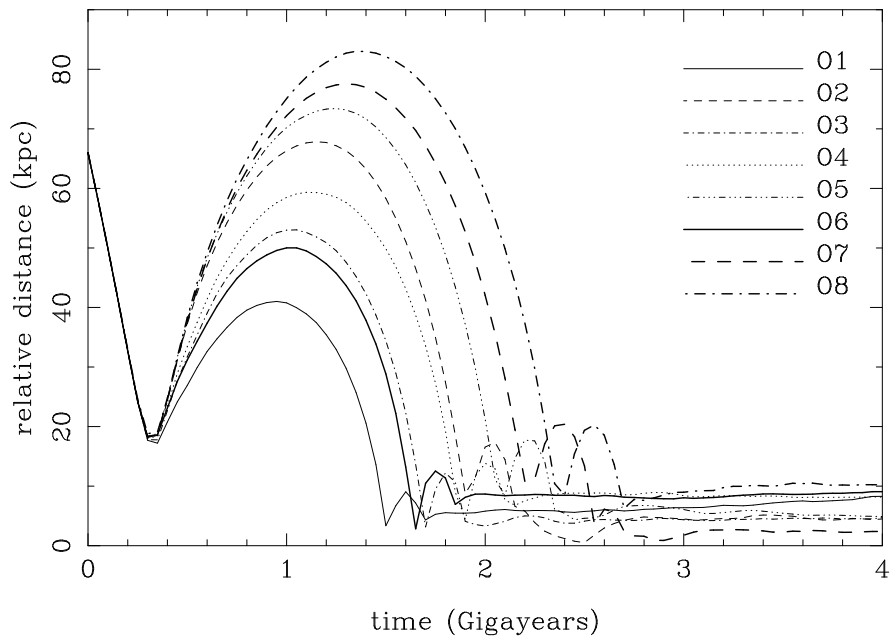}
\vspace{9cm}
\caption{Time evolution of the relative distance of the two galaxies for each
of the relative orientations.}
\end{figure}

During the merging of both galaxies some particles are lost, specially in the
cases of direct encounters. In the case of perpendicular orientations or 
bigger spin differences the fraction of loose particles is very small 
(less than 1\%). For parallel spins the fraction of particles lost amounts to
4\% while in the case of spin difference of 45$^{\circ}$ the amount is the
3\%. We also note that in direct interactions the discs
loose their particles at the first passage, while for the other orientations
the small fraction of particles is lost at the moment of the merging.

\end{document}